# Recent advances, perspectives and challenges in ferroelectric synapses

Bobo Tian[1,2,3]* Ni Zhong[1] and Chungang Duan[1,4]*

[1]Key Laboratory of Polar Materials and Devices (MOE), Department of Electronics, East China Normal University, Shanghai 200241, China

[2]State Key Laboratory of Infrared Physics, Shanghai Institute of Technical Physics, Chinese Academy of Sciences, 500 Yu Tian Road, Shanghai 200083, China.

[3]Laboratoire Structures, Propriétés et Modélisation des Solides, CentraleSupélec, CNRS-UMR8580, Université Paris-Saclay, 91190 Gif-sur-Yvette, France.

[4]Collaborative Innovation Center of Extreme Optics, Shanxi University, Shanxi 030006, China

*Correspondence: bbtian@ee.ecnu.edu.cn and cgduan@clpm.ecnu.edu.cn

Abstract

The multiple ferroelectric polarization tuned by external electric field could be used to simulate the biological synaptic weight. Ferroelectric synaptic devices have two advantages compared with other reported ones: One is the intrinsic switching of ferroelectric domains without invoking of defect migration as in resistive oxides contributes reliable performance in these ferroelectric synapses. Another tremendous advantage is the extremely low energy consumption because the ferroelectric polarization is manipulated by electric field which eliminate the Joule heating by current as in magnetic and phase change memory. Ferroelectric synapses are potential for the construction of low-energy and effective brain-like intelligent networks. Here we summarize recent pioneering work of ferroelectric synapses involving the structure of ferroelectric tunnel junctions (FTJs), ferroelectric diodes (FDs) and ferroelectric field effect transistors (FeFETs), respectively, and shed light on future work needed to accelerate their application for efficient neural network.

## 1. Introduction

The constant data shuttling between the central processing unit (CPU) and memory units, inevitably results in considerable latency and large additional power consumption in the von Neumann architecture. This problem becomes more severe for intelligent tasks, such as computer vision, speech recognition and autonomous vehicles, which imposes critical requirements on the energy efficiency and processing speed. Modified architectures with enhanced computing capability and efficiency are urgently expected. The human brain with a network structure consisting of neurons interconnected by synapses, can perform massive parallel operating with extremely low energy (~1-10 fJ per event)[1]. A great amount of efforts is devoted to imitate the efficiency of the human brain, and among which neural network hardware is the most straightforward and effective method. The connection strength of biological synapses can be reconfigured in a plastic manner by electrical signals from the pre- and post-neurons. The synaptic connection strength is typically emulated by the electronic conductance for neural network hardware. Therefore, a continuous conductance modification and similar plasticity is essential for artificial synapses. Depending on the application of neural networks, well nonvolatility is usually also needed in the artificial synapses. Memristor is such kind of device and is usually designed for achieving synaptic operations such as short- and long-term plasticity and spike-timing dependent plasticity (STDP) and for accomplishing recognition or other complicated tasks when they are connected in artificial neural networks (ANN).

Various materials, such as filament-forming oxides, 2D materials, phase change materials, magnetics and ferroelectrics, have been proposed for well performed memristors[2-4]. Among these materials, ferroelectrics are promising candidates for mimicking artificial synaptic devices and performing neural network operations because of their nonvolatile multilevel memory effect. The spontaneous electric polarization in the ferroelectrics can be reversed step by step under an external electric field. Both theoretical calculations[5] and direct experimental observations[6] confirm the uninterrupted ferroelectric domain dynamics. By controlling the amplitude or duration of the applied electric field pulse, continuous intermediate states of ferroelectric domains can be obtained (Figure 1). These intermediate domain states are non-volatile and may maintain as long as 10 years depending on the electrode screening. It is worth noting that the continuous evolution of ferroelectric domains under external field is very similar to the update of synaptic connection weight in biological brain learning. Thus, the continuous evolution of ferroelectric domains is treated as "ferroelectric plasticity" and is used to design ferroelectric synapses for mimicking biological synaptic plasticity.

Ferroelectric tunnel junctions (FTJs), ferroelectric diodes (FDs) and ferroelectric field effect transistors (FeFETs) are promising candidates due to their ferroelectric associated resistive switching effect. The ferro-resistive switching behavior is based on the intrinsic switching of ferroelectric domains without invoking of defect migration as in resistive oxides. Therefore, ferroelectric synapses possess a fundamental merit over defect-involved resistive synapses for achieving reliable performance once reliable fabrication parameters are established. Another tremendous advantage of these artificial ferroelectric synaptic devices is the extremely low energy consumption because the ferroelectric polarization is manipulated by electric field which eliminate the Joule heating by current as in magnetic and phase change memory.

Furthermore, emerging CMOS-compatible ferroelectric materials, such as hafnium oxide ($HfO_2$) and organic polymers, make it meet the highest salability state of art. Thus, ferroelectric synapses are potential for the construction of effective brain-like intelligent networks. Here we summarize recent pioneering work of ferroelectric synapses involving the structure of FTJs, FDs and FeFETs, respectively, and shed light on future work needed to accelerate their application for efficient neural network.

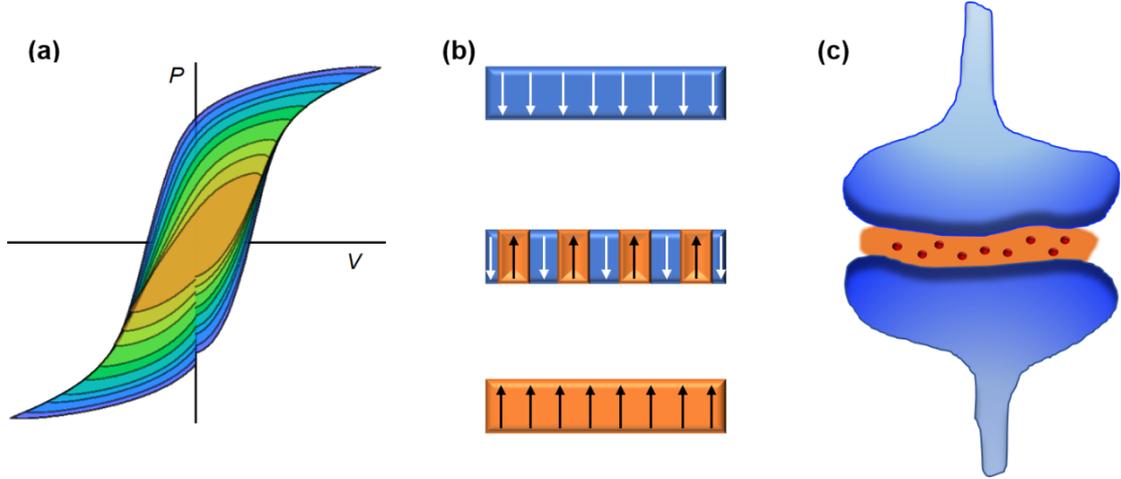

Figure 1 (a) Typical multiple polarization (*P*) versus voltage loops under various voltage amplitude obtained from an Au/PVDF/Al capacitor. (b) Schematic illustration of multiple domains states in a ferroelectric film. (c) Sketch of a synapse.

## 2. Ferroelectric synapses based on FTJs
### 2.1 History of FTJs

FTJs are two-terminal devices in which an ultrathin ferroelectric layer is sandwiched between two electrodes. The thickness of this ferroelectric layer should be only a few unit-cells for quantum-mechanical electron tunneling. As Figure 2 shows, the polarization switching alters the barrier via electrostatic effects at the ferroelectric/electrode interfaces, resulting in a giant tunnel electroresistance (TER)[7].

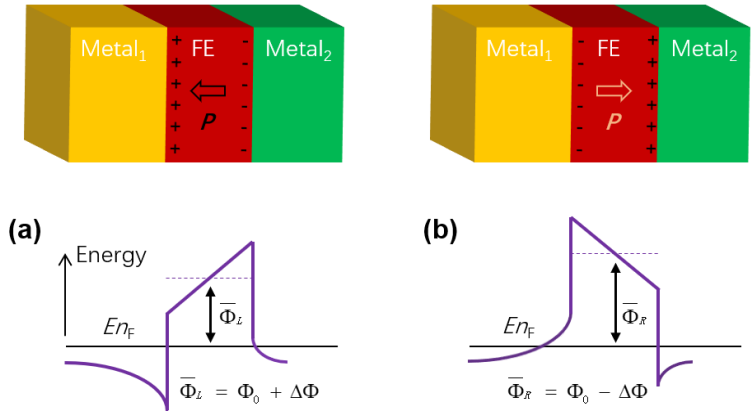

Figure 2 (a-b) Schematic representation of the energy potential profile in an FTJ for polarization pointing to the electrode one ($M_1$) (a) and for polarization pointing to the electrode two ($M_2$) (b). Reproduced from Ref. [7], Copyright 2005, with permission of APS.

The concept of FTJs was proposed by Esaki in 1971 (ref. [8]) and drew experimental attention until the realization of ultrathin inorganic ferroelectric films 30 years later[9-11]. In 2009, three groups independently demonstrated TER effect triggered by polarization reversal on bare surfaces of ferroelectrics[12-14]. Following this joint discovery, active research work is devoted to solid state FTJs. In 2011, Pantel *et al*.[15] fabricated submicron devices based on 9 nm $PbZr_{0.2}Ti_{0.8}O_3$ sandwiched between $La_{0.7}Sr_{0.3}MnO_3$ bottom electrodes and Cu top electrodes and reported the resistive switching accompanied by ferroelectric polarization reversal. In 2012, Chanthbouala *et al*.[16] demonstrated the TER effect triggered by polarization reversal in the tunnel junctions based 2-nm $BaTiO_3$ sandwiched between $La_{0.7}Sr_{0.3}MnO_3$ bottom electrodes and Au/Co top electrodes. After this, solid-state FTJs based on ultrathin $BaTiO_3$ (refs[17-22]), $PbZr_{0.2}Ti_{0.8}O_3$ (ref.[23]) and $BiFeO_3$ (refs[24]) films are reported extensively. Various efforts such as strain[24], electrode/barrier interface engineering[18, 20] and functional electrode[17] have been introduced to increase the TER OFF/ON ratio in these oxide FTJs[25]. The TER OFF/ON ratio can be as large as $10^7$ at room temperature[26]. In addition, using ferromagnetic metals as electrodes, enable a nonvolatile control of electronic and spintronic responses in multifunctional devices[27-31]. It is a remarkable fact that CMOS-compatible ferroelectric materials, such as hafnium oxide[32-36] and organic polymers[37], are recently adapted to FTJs, make it closer to commercialization.

### 2.2 Inorganic FTJ synapse

Apart from the binary up and down states of the spontaneous electric polarization in ferroelectrics, the uninterrupted dynamics allow abundant intermediate states of ferroelectric domains. The spontaneous electric polarization can be reversed step by step under external electric field pulses by altering their amplitude and duration. These intermediate domain states are non-volatile and may maintain as long as 10 years depending on the electrode screening. The continuous evolution of ferroelectric domains is very similar to the update of synaptic connection weight in biological brain learning. By modulating domain dynamics in the ferroelectric barrier, the junctions achieve memristive behaviour[38-40] and are used as ferroelectric synapses for brain-inspired computation[36, 41-46].

The evolution of the junction resistance and ferroelectric domain configuration are demonstrated in $BaTiO_3$ and $BiFeO_3$ FTJs (Figure 3a)[24, 38]. The progressive resistance evolution accompanied by the reversal of the ferroelectric polarization could be realized not only by altering the applied voltage amplitude, but also by an appropriate number of consecutive pulses with a fixed voltage. The analog response of the junctions could be illuminated by a parallel conductance model (Figure 3b), in which the reversal of ferroelectric domains is dominated by nucleation mechanism. This memristive behavior is appealing for artificial synapses, where synaptic weights are usually simulated by value of device conductance.

The update of synaptic strengths contributes learning in spiking neural networks (SNN). One of particularly promising learning rule is spike timing-dependent plasticity (STDP) through which the synaptic weights evolve depending on the timing and causality of electrical signals from neighboring neurons[47, 48]. As sketched in Figures 4a and 4b, morphologically, the FTJ could be treated as a two-terminal synapse where the top and bottom electrodes transmit signals from pre and post neuron. In the hysteresis cycle of the resistance versus voltage amplitude, there are voltage thresholds $\pm V_{th}$ beyond which resistive switching occurs (Figure 4c). This well-defined voltage thresholds are associated with the coercivity of the ferroelectric. By designing shape of pre and post voltage signals where only a superposition of ($V_{pre}$ - $V_{post}$) transitorily exceeds the threshold voltage (inset of Figure 4d), Boyn et al. demonstrated the STDP learning rule in $BiFeO_3$ FTJs where the junction conductance increases or decreases depending on the timing and causality of pre and post electrical signals (Figure 4d)[43].

Ma et al. fabricated Nb doped SrTiO$_3$ (Nb:SrTiO$_3$) electrode based FTJs in which the Nb:SrTiO$_3$ electrode with a higher carrier concentration and a metal electrode with lower work function was reported could improve the operation speed[49]. 32 conductance states and STDP learning rule with ultrafast operation are obtained in their Ag/BaTiO$_3$/Nb:SrTiO$_3$ FTJs. The fastest operation speed is of 600 ps and maintains up to 358 K. These results imply the potential of FTJs for developing ultrafast neuromorphic computing systems.

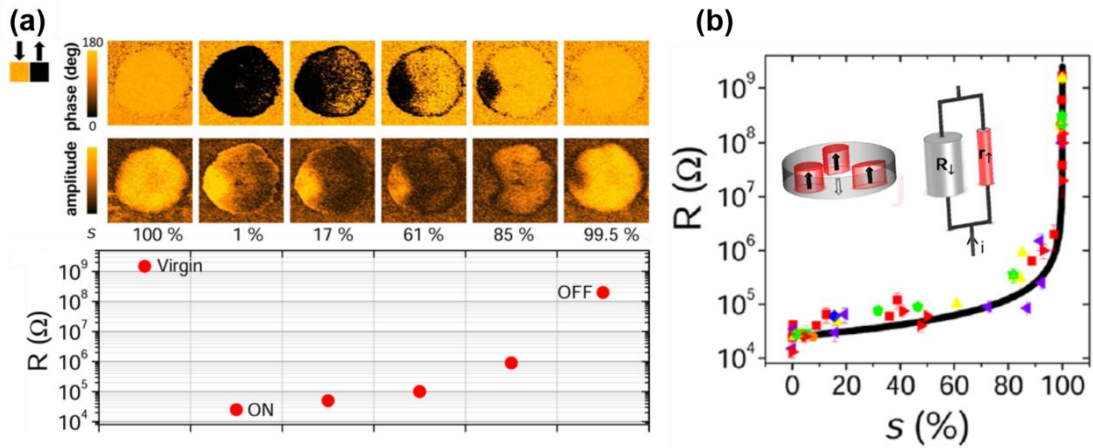

Figure 3 (a) Evolution of the junction resistance and ferroelectric domain configuration in a 180 nm wide (Ca,Ce)MnO$_3$/BiFeO$_3$/Pt/Co FTJ. (b) Symbols are experimental resistance as a function of percentage of down domains extracted from the PFM phase images and the black curve is a simulation in a parallel resistance model shown in inset. Reproduced from Ref. [24], Copyright 2013, with permission of ACS.

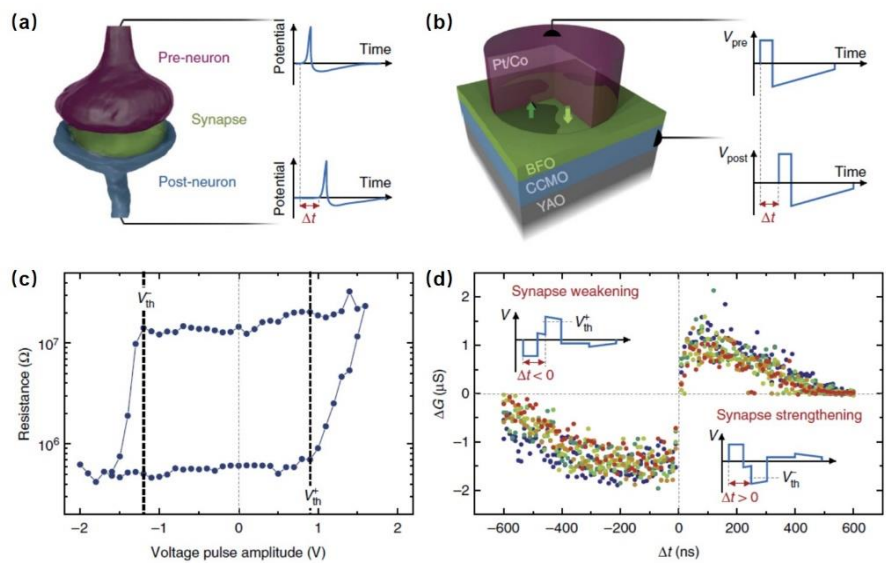

Figure 4 The FTJ could be treated as a two-terminal synapse where the top and bottom electrodes transmit signals from pre and post neuron. (a) Sketch of pre- and post-neurons connected by a synapse. (b) Sketch of the (Ca,Ce)MnO$_3$/BiFeO$_3$/Pt/Co FTJ. YAO stands for YAlO$_3$. (c) Single-pulse hysteresis loop of the resistance versus

voltage amplitude, there are voltage thresholds ±$V_{th}$ beyond which resistive switching occurs. (d) Measurements of STDP in the FTJ. The junction conductance increases or decreases depending on the timing and causality of pre and post electrical signals. The insets show the waveforms produced by the superposition of presynaptic and postsynaptic spikes. Reproduced from Ref. [43], Copyright 2014, with permission of Springer Nature.

Using ferromagnetic metals as electrodes introduce a new spintronic control in FTJs. Huang et al. fabricated a $La_{0.7}Sr_{0.3}MnO_3$/$BaTiO_3$/$La_{0.7}Sr_{0.3}MnO_3$ multiferroic tunnel junction (MTJ) and demonstrated that the ferroelectric domain dynamics characteristics, thus the synaptic plasticity could be manipulated by the relative magnetization alignment of the electrodes[44]. As shown in Figure 5, the STDP is altered by changing the magnetization alignment of the electrodes between parallel (P) or antiparallel (AP) states. The small value of difference may be due to the weaker magnetoresistance effect compared with the large electroresistance effect by altering the ferroelectric domains in MTJs. It is expected that the strength of manipulation for synaptic plasticity could be further enhanced when the spin parameter contributes memristive behaviors and ferroelectric domians is used as modulation in MTJs. The multiple and controllable plasticity characteristic in a single artificial synapse using MTJs is conducive to the development of artificial intelligence, thus could rise the enthusiasm of research region for magnetoelectric coupling.

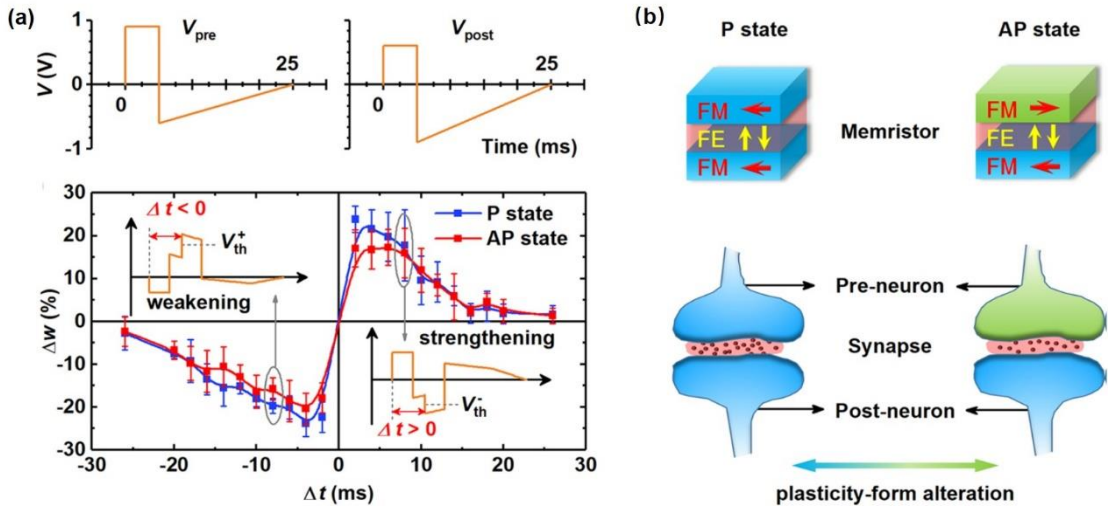

Figure 5 (a) Measurements of STDP in $La_{0.7}Sr_{0.3}MnO_3$/$BaTiO_3$/$La_{0.7}Sr_{0.3}MnO_3$ MTJ with the parallel (P) or antiparallel (AP) states of magnetization alignment in electrodes. (b) Schematic illustration of the magnetoelectrically coupled memristor and the corresponding synapses. Reproduced from Ref. [44], Copyright 2018, with permission of ACS.

### 2.3 Organic FTJs synapses

Sophisticated experimental approaches such as strain engineering and careful control of epitaxial growth are required for ferroelectricity in nanometer-thick perovskite films, which inevitably results in a complex fabrication process[50]. On the route toward non-volatile memories or ANN combining ferroelectric devices with CMOS technology, organic ferroelectric polymers have also been considered as tunnel barriers. Ferroelectric poly(vinylidene fluoride) (PVDF)[51-53] and its copolymers, P(VDF-TrFE)[54-56], maintain a robust ferroelectricity with a remnant polarization of ~10 μC/cm$^{-2}$ at room temperature and be processed from solution on practically any material, potential for flexible device integration and large-scale fabrication[57]. The weak Van der Waals interfacial bonding with metal electrodes in organic FTJs may introduce more controllable electronic transport properties than their inorganic counterparts.

Using high-sensitivity piezoresponse force microscopy (PFM), Tian et al. demonstrated that the ferroelectric polarization remains switchable with low voltages in films as thin as one or two layers of ferroelectric poly(vinylidene fluoride) (Figure 6a)[37]. Based on these high-quality ferroelectric polymer films as tunnel barrier, they fabricated submicron organic Au/PVDF/W FTJs on silicon substrates (Figure 6b), in which the resistive switching company with polarization reversal and the TER reaches 1000% at room temperature (Figures 6c-e). The TER is quantitatively explained by electrostatic effects in a direct tunnelling regime. The TER could be further enhanced to $10^7$% by inducing a semiconducting Nb:SrTiO$_3$ electrode[58].

Majumdar et al. demonstrated the synapse behaviour in the Au/P(VDF-TrFE)/Nb:SrTiO$_3$ FTJs[40]. By altering the applied voltage amplitude, the organic FTJ shows multi progressive resistance evolution loops (Figure 7a). Programmable the conductance, corresponding to LTP/D synaptic weight, could be achieved by modulating either amplitude (Figure 7b) or number (Figure 7c) of the applied voltage pulses. Other synaptic functions like short term potentiation and depression (Figure 7d), paired-pulse facilitation and depression, and Hebbian and anti-Hebbian learning rule ((Figure 7e) are also demonstrated in these organic FTJs. These results shine light on an outlook that organic FTJs could be used as building blocks in ANN.

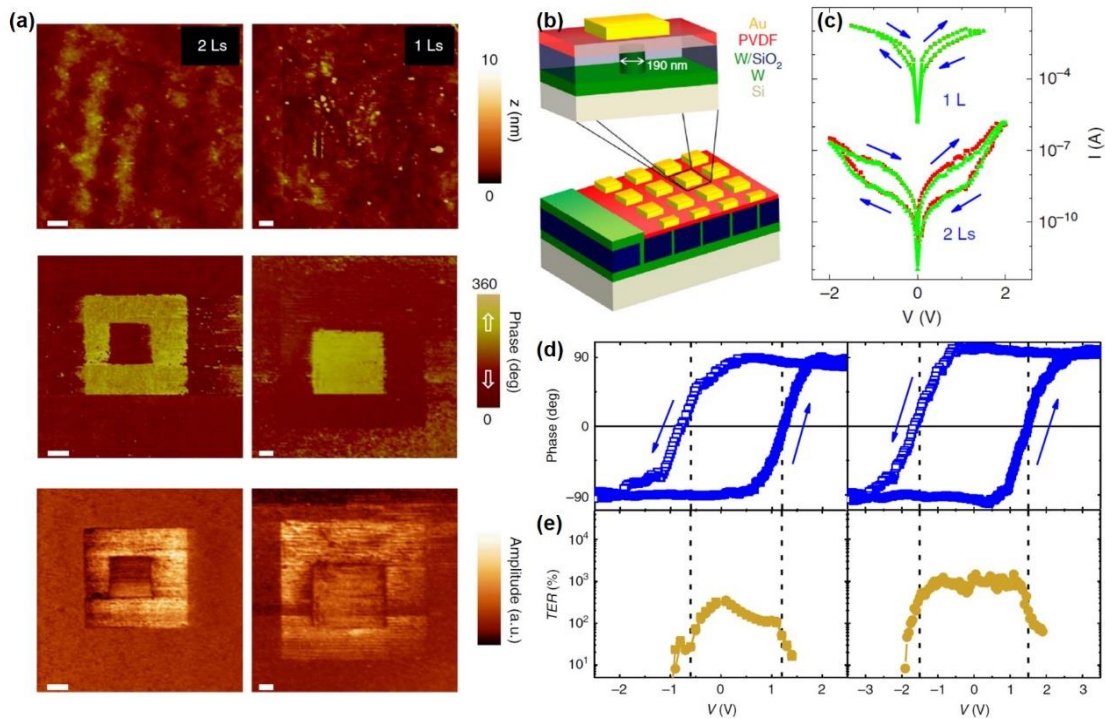

Figure 6 (a) Ferroelectricity in ultrathin thin films of PVDF. Topography (top), PFM phase (middle) and amplitude (bottom) images, after switching the polarisation upward in the central 5x5 μm$^2$ area of a 2Ls-thick (left) and 1L-thick (right) PVDF films. Scale bar, 1 mm. (b) Three-dimensional schematic diagram of PVDF FTJs. (c) *I-V* curves in the 1 and 2Ls PVDF FTJs. The arrows show the direction of the voltage sweeps. (d) Local PFM phase and (e) Voltage dependences of the TER of an Au/PVDF (1L)/W junction (left) and an Au/PVDF (2L)/W junction (right), respectively. The vertical dashed lines show the agreement between polarisation and resistance coercive voltages. Reproduced from Ref. [37], Copyright 2016, with permission of Springer Nature.

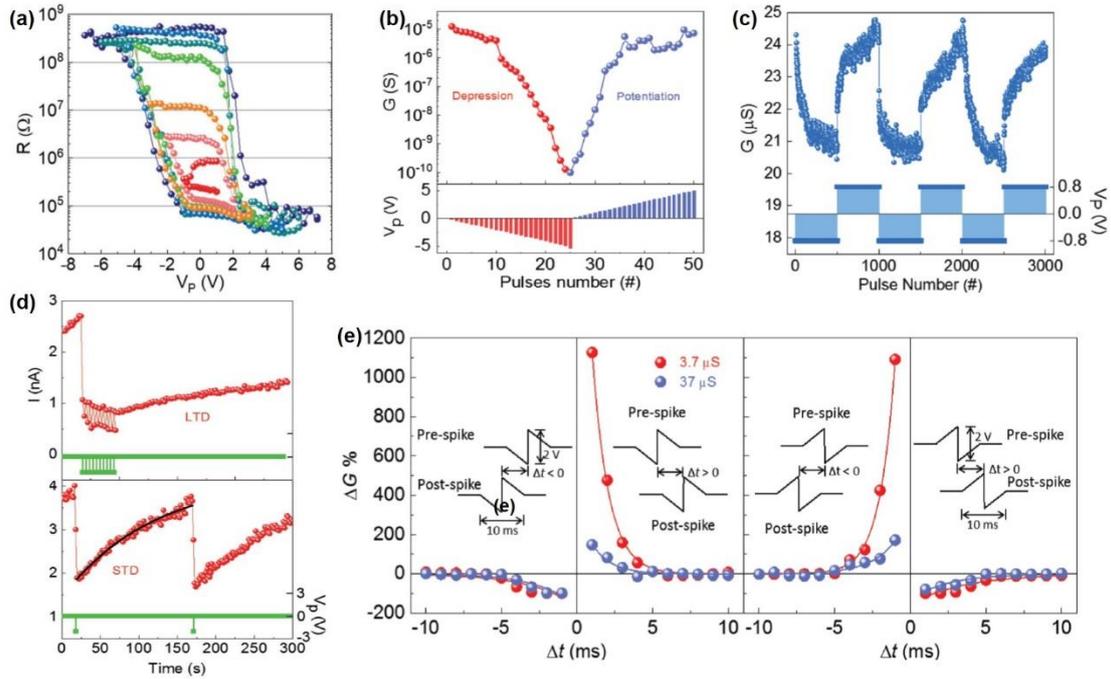

Figure 7 (a) Evolution of the junction resistance as a function of voltage amplitude in an Au/P(VDF-TrFE)/Nb:SrTiO$_3$ FTJ. (c-d) Programmable the conductance, corresponding to LTP/D of synaptic weight, is achieved by modulating either amplitude (b) or number (c) of applied voltage pulses. (d) Measurements of LTP/D. (e) Measurements of Hebbian and anti-Hebbian STDP learning rule. Reproduced from Ref. [40], Copyright 2019, with permission of John Wiley and Sons.

**2.4 Design of FTJs synapses array**

The memristor crossbar array supports high density integration and parallel operation, which can mimic a neural network to perform data intensive applications such as machine learning tasks[2, 59]. The online training of the neural network requires linearity and symmetry of conductance update in memristors[60]. It asks that a single programming pulse should change the weight by a constant value so that the calculated weight update (*Δ*w) should be directly transformed to the number of programming pulses for the memristor, without extra checking and accounting operations in the training process.

Unlike the filament-based memristors suffering from nonlinear and asymmetric conductance update due to the intrinsic non-linear dependence of conductance on filament shape[2], FTJs synapses are reported to achieve high conductance update linearity and symmetry with small switching variations[45]. The analog response of FTJs follows a parallel conductance model in which the reversal of ferroelectric domains is dominated by nucleation mechanism. For example, the conductance is small for up polarization in Pt(top)/BTO$_3$/Nb:SrTiO$_3$(bottom) FTJs (Figure 8a). The intermediate conductance states ($G_m$) could be described by small conductance with up polarization ($G_{ON}$), large conductance with down polarization ($G_{OFF}$) and normalized down polarized area ($S_{ON}$): $G_m$ = (1-$S_{ON}$)x$G_{OFF}$ + $S_{ON}$x$G_{ON}$ = ($G_{ON}$-$G_{OFF}$)x$S_{ON}$ + $G_{OFF}$. The conductance exhibits a linear relation with the ferroelectric domain evolution. By finely tuning the applied voltage pulse parameters, Li et al. realized gradual ferroelectric domain switching and obtained linear and symmetric conductance update with 200 states and 500 reproducible cycles (Figure 8b)[45]. The switching linearity is the best one among reported two-terminal electronic synapses. A simulated ANN consisted of three layers is built from two FTJs arrays (Figure 8c), supervised learning with backpropagation algorithm exhibited very high learning accuracies of 96.5% for 8 × 8 pixel image version (small image) of handwritten digits from the University of California at Irvine (UCI) image data set and 96.4% for 28 × 28 pixel image version (large image) of handwritten digits from the Mixed National Institute of Standards and Technology (MNIST) data set (Figure 8d).

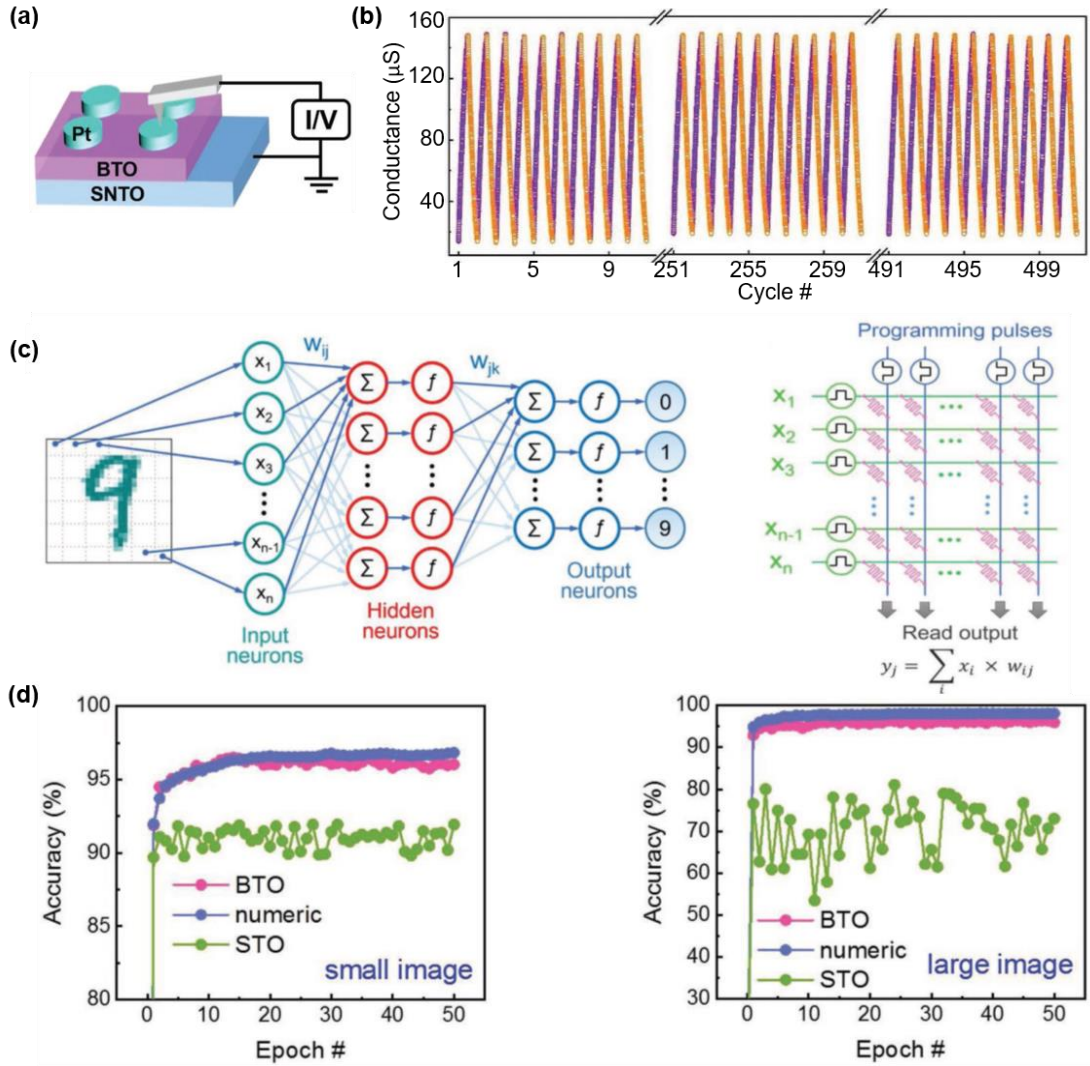

Figure 8 (a) Sketch of Pt/BaTiO$_3$/Nb:SrTiO$_3$ FTJs. (b) Linear and symmetric conductance update with 200 states and 500 reproducible cycles. (c) Schematic diagram of a three layers neural network (left) and neural core with a crossbar structure to perform the analog matrix operations (right). (d) The simulation results, the training accuracy of BaTiO$_3$ FTJs and SrTiO$_3$ based resistive devices for a small image (left) and a large image (right). Reproduced from Ref. [45], Copyright 2019, with permission of John Wiley and Sons.

Recently, the CMOS-compatible fluorite-structure HfO$_2$, enable low-temperature synthesis and directly growth on silicon, is reported to maintain a ferroelectric polarization by appropriate doping, such as Hf$_{0.5}$Zr$_{0.5}$O$_2$ (HZO) thin films[61], which enhance the possibilities to integrate ferroelectric films into CMOS processes. Since the discovery of ferroelectricity in HfO$_2$-based thin films in 2011[62], various factors including stress[63], oxygen vacancies[64] and surface/interface energy[61] were discussed for the cause of the ferroelectric properties in HfO$_2$-based thin films. There are various crystalline phases in HfO$_2$-based films, in which the monoclinic (M−) phase for hafnium-rich mixed HZO films and the tetragonal (T-) phase for high ZrO$_2$ content are non-polar structure, while the orthorhombic (O−) Pbc2$_1$ phase for Hf/Zr ratio of around 0.5 is polar structure[64]. The polar 'rhombic' distortion (2c/(a+b)) is due to the centre anion (oxygen) displacement relative to its surrounding cation tetrahedron (Hf or Zr) in the polar O-phase, while the oxygen lies in the polyhedral centre of the tetrahedron in the nonpolar T-phase. The thick or bulk HZO is in the non-polar M-phase, with the decrease of the thickness, high-symmetry non-polar T-phase and polar O-phase are preferred due to the size effects that surface energies favors higher symmetry. The symmetry in O-phase is lower than that in T-phase and distortions favors lower symmetry. Special electrode materials and a rapid post-metal annealing process are usually used to induce confinement strain, thus distortion, to the films, which stabilizes the desired polar orthorhombic phase. On the other words, coupling intrinsic (surface energies) and extrinsic (confinement strain) mechanisms can favor inversion symmetry breaking in ultrathin HZO films. The robust ferroelectricity in HZO films of thickness down to 1 nm is reported[62].

Quantity of $HfO_2$-based FTJs spring up[32-36]. The CMOS-compatible $HfO_2$-based FTJs are considered as practical candidate of ferroelectric synapses[36, 65, 66]. Chen et al. fabricated a three-dimensional vertical HZO-based FTJs with a remnant polarization of 16 $\mu C/cm^{-2}$ and fast non-volatile storage at 100 ns for over $10^3$ cycles[36] (Figure 9a). The comprehensive electronic synaptic functions including long-term potentiation and depression (LTP/D) and Hebbian and anti-Hebbian STDP learning rule (Figure 9b) were achieved in their HZO-based FTJs. Pattern training with strong tolerance to input faults and variations is implemented in their 3D vertical FTJs array. Furthermore, simulation of a two-layers network with two HZO-based FTJs arrays (Figure 9c) demonstrate that accuracy rate of the pattern classification and recognition is higher than 96%. These results demonstrate a high potential of HZO-based FTJs to form complex hardware neural network applications.

Despite the pioneering synaptic properties such as nonvolatility, gradual conductance change, threshold feature, energy efficiency, high uniformity and simple structure in single FTJ device, there is no true hardware neural network based on FTJs synapses up to now. Obstacles include the strict epitaxy and high temperature requirements during the fabrication of perovskite FTJs and the instability of ferroelectricity in hafnium-based oxide such as stain requirement and poor endurance.

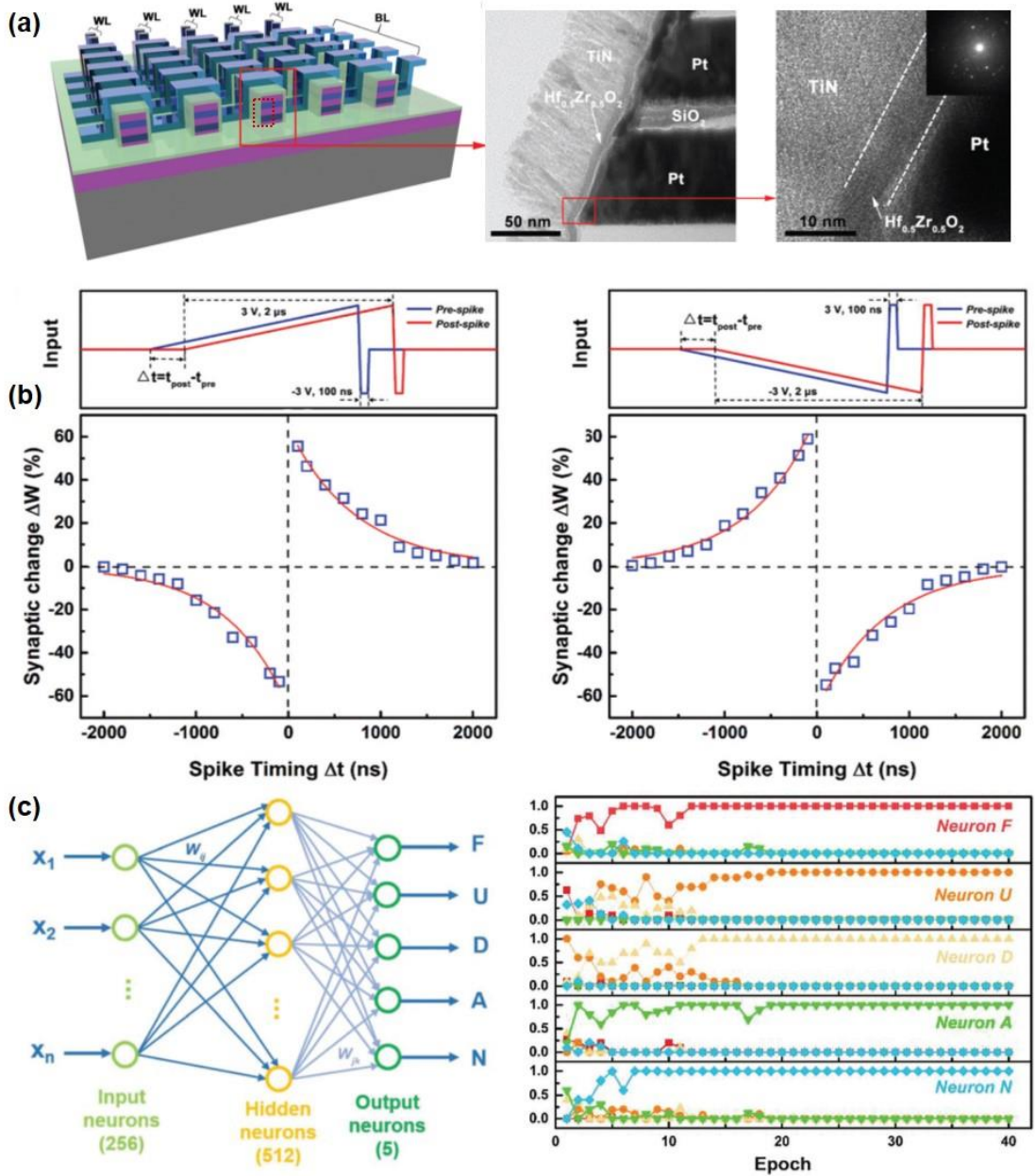

Figure 9 (a)The schematic diagram of high-density 3D vertical HZO-based FTJS array with high resolution transmission electron microscopy image of the 3D TiN/ HZO/Pt devices. (b) Measurements of Hebbian and anti-Hebbian STDP rule in the HZO-based synaptic array. The red lines are the exponential fits to the experimental data. (c) The schematic diagram of multilayer perceptron network (left), and simulated output signal in which 'correct' class of the applied letter patterns is larger than the other 'false' outputs after learning process (right). Reproduced from Ref. [36], Copyright 2018, with permission of RSC.

## 3. Ferroelectric synapses based on FDs

Endurance in FTJs has been demonstrated only up to a few $10^8$ cycles[49], this is further inferior compared with the capacitive ferroelectric memories (> $10^{16}$ cycles). The degeneration of ferroelectricity may result from the pinning effect of domain walls due to rearrangement of defects[50]. This pinning effect is much stronger in nanometer-thick ferroelectric films of FTJs. FDs have the similar structure with FTJs but the ferroelectric barrier is much thicker and the resistive switching is dominated by one of the ferroelectric/electrode interface[67].

The mechanism how does the polarization control the resistive switching behavior is disputable. The mostly proposed view is that the interfacial depletion-layer width is modulated by ferroelectric polarization reversal, resulting in two resistive states in FDs. Figure 10 shows schematic diagram of charge distribution and energy-band diagrams of a Pt/BiFeO$_3$/Nb:SrTiO$_3$ FD at low resistance state (LRS) and high resistance state (HRS), respectively[68]. Based on the electrostatic theory, when ferroelectric polarization points to the interface, free electrons will be attracted to screen the positive polarization charges, resulting in a reduced depletion-layer width. In this case, the FD is on the LRS (Figure 10a). When ferroelectric polarization points away from the interface, holes will be depleted by the negative polarization charges, leading to an enhanced depletion layer width. Then the FD is on the HRS (Figure 10b).

Various resistive FDs, such as based on BiFeO$_3$[68-75], PbTiO$_3$[67], BaTiO$_3$[76] and so on, were widely researched. One can find more information in a recent review focusing on ferroelectric heterostructures[77]. Below shines a light on how do FDs are used as synaptic devices.

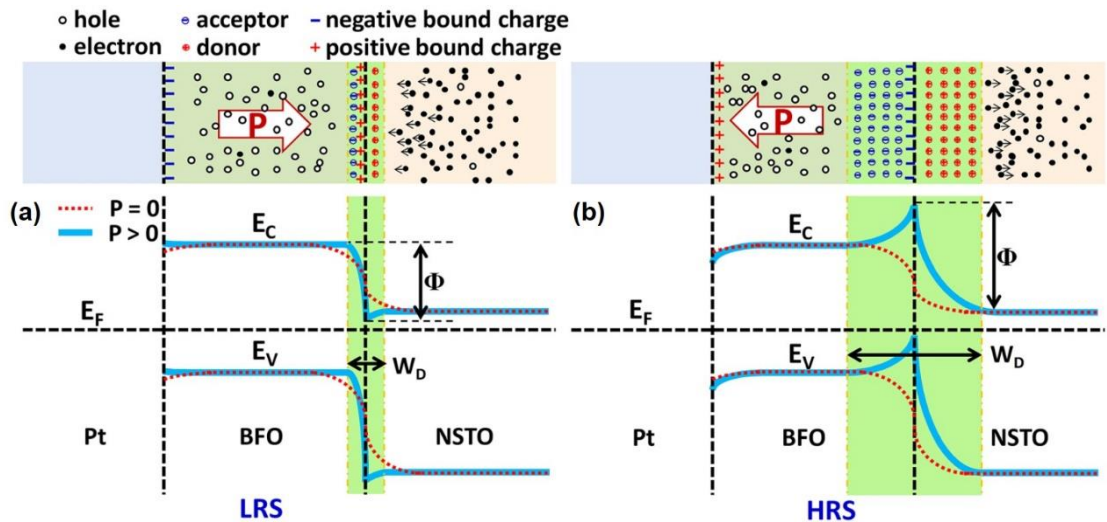

Figure 10 (a) Schematic diagram of charge distribution and energy-band diagrams of a Pt/BiFeO$_3$/ Nb:SrTiO$_3$ FD at low resistance state (LRS). (b) Schematic diagram of charge distribution and energy-band diagrams of a Pt/BiFeO$_3$/ Nb:SrTiO$_3$ FD at high resistance state (HRS). Reproduced from Ref. [68], Copyright 2013, with permission of AIP Publishing.

Jia et al. demonstrated the correlation between the resistive switching and polarization reversal in Au/BaTiO$_3$/SmNiO$_3$ FDs[76]. Figure 11a shows the configuration of the measurements. Resistive switching with resistance ratios greater than 100 were observed in both off-pulse $R$-$V_{write}$ (resistance versus write voltage) curves (Figure 11b) and DC $I$-$V$ curves (Figure 11c). The threshold voltage of resistive switching corresponds well with the coercive voltage of ferroelectric BaTiO$_3$ as demonstrated in the PFM hysteresis loops (Figure 11d). The resistance switching between polarization up and down is experimentally observed using a conductive PFM technique in BiFeO$_3$ FDs (Figure 12a). These results imply that the resistive switching is most probably dominated by ferroelectric domains in FDs. By modulating the ferroelectric domains, the FDs are alternate ferroelectric synaptic devices.

The memristive behaviour in BiFeO$_3$ FDs is robust at ultra-wide temperature range from -170 ℃ to 300 ℃[69]. The ferroelectric BiFeO$_3$ have an extremely high Curie temperature of ~930 ℃, which insure its ferroelectricity at temperature as high as 500 ℃ (Figure 12b). The analogy resistive switching with identical threshold (Figure 12b) and synaptic functions such as Hebbian STDP learning rule (Figure 12c) are demonstrated in Au/BiFeO$_3$/SrRuO$_3$ FDs at temperature range from -170 ℃ to 300 ℃. These results show that ferroelectric synapses provide a feasible way for the realization of neural network in extremely wide temperature applications.

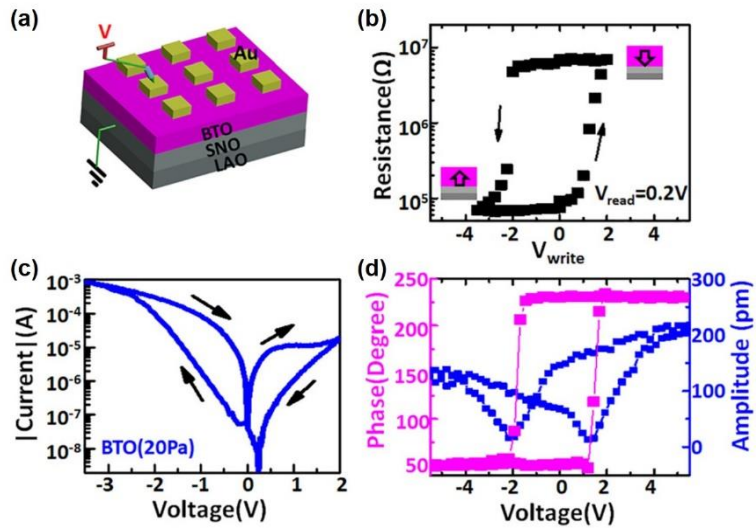

Figure 11 (a) Schematic diagram of Au/BaTiO$_3$/SmNiO$_3$ FDs. (b) Off-pulse $R$-$V_{write}$ curves in an Au/BaTiO$_3$/SmNiO$_3$ FD. (c) DC $I$-$V$ curves in an Au/BaTiO$_3$/SmNiO$_3$ FD. (d) Local PFM hysteresis loops of an Au/BaTiO$_3$/SmNiO$_3$ FD. Reproduced from Ref. [76], Copyright 2019, with permission of AIP Publishing.

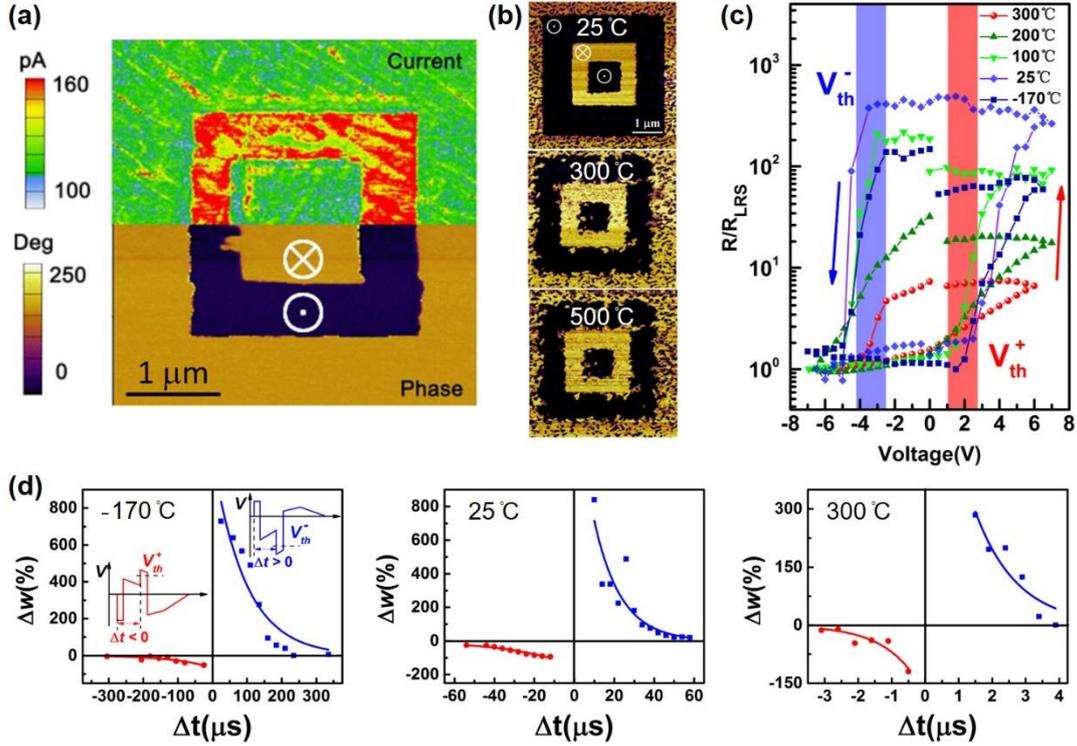

Figure 12 (a) The lower half part is out-of-plane phase image and the upper half part is the corresponding current mapping with read voltage of -0.5 V measured on $BiFeO_3/SrRuO_3/SrTiO_3$ (substrate). (b) Out-of-plane phase image after heating the sample at 25 ℃, 300 ℃ and 500 ℃ for 5 min, respectively. (c) *R-V* hysteresis loops of an $Au/BiFeO_3/SrRuO_3$ FD under -170 ℃, 25 ℃, 100 ℃, 200 ℃ and 300 ℃, respectively. Reproduced from Ref. [69], Copyright 2019, with permission of IOP Publishing.

## 4. Ferroelectric synapses based on FeFETs

In a FeFET device, the ferroelectric domains in gate tune the carrier concentration of semiconductor channel layer, and hence programs multiple channel conduction states, making FeFET are ideal for artificial synapses. Note that the input port can be either the gate-source/drain port or the drain-source port. For example, for supervised learning in general SNN, the channel usually connects pre and post synaptic neurons while the gate terminal accepts signals to modulate the channel conductance. Compared to two-terminal synapse devices, this three-terminal structure has an advantage of direct feedback modulation to synapse weight from the gate electrode, which allows concurrent learning. Some typical experimental results of the FeFET synapses are summarized in Table 1.

Table 1 Experimental results of FeFET synapses.

| Ferroelectric gate insulator | Channel | Ratio | States | Retention | Energy | Synaptic Functions | Year | Ref. |
|---|---|---|---|---|---|---|---|---|
| $SrBi_2Ta_2O_9$ | Si | 2 | 8 | - | - | - | 1999 | [78] |
| $Pb(Zr,Ti)O_3$ | ZnO | >$10^5$ | >10 | >$10^5$ s | - | LTP/D、STDP、On-chip pattern recognition | 2014 | [79] |
| HZO | Si | 45 | 32 | - | - | LTP/D、Simulation of image recognition | 2017 | [80] |
| Si:HfO$_2$ | Si | 3 | 32 | - | - | LTP/D、STDP | 2017 | [81] |
| HZO | Si | 4.98 | >32 | | | LTP/D、Simulation of pattern recognition | 2018 | [82] |
| HZO | IGZO | 14.4 | 65 | - | - | LTP/D、Simulation of pattern recognition | 2019 | [83] |
| P(VDF-TrFE) | Organic polymer | $10^4$ | - | >$10^4$ s | 75 PJ | STP*、LTP、EPSC#、Visual-perception | 2018 | [84] |
| P(VDF-TrFE) | MoS$_2$ | $10^4$ | 1000 | >$10^3$ s | <1 fJ | LTP/D、STDP | 2019 | [1] |
| P(VDF-TrFE) | Graphene | 3.2 | 50 | - | 50 nJ | Complementary synapses、Simulation of pattern classification | 2019 | [85] |
| P(VDF-TrFE) | Organic polymer | 5 | 8 | - | - | - | 2019 | [86] |
| P(VDF-TrFE) | Carbon-nanotube | $10^4$ | 100 | 400 s | - | EPSC、IPSC&、LTP/D、Simulation of image recognition | 2020 | [87] |

*STP: short term potentiation; # EPSC: excitatory postsynaptic currents; *STP: inhibitory postsynaptic currents.

**4.1 Inorganic FeFET synapses**

In 1993, Ishiwara proposed that the analog memory behavior in FeFETs could be used in adaptive-learning neuron circuits, in which the channel conductance of FeFETs is treated as synaptic weights[88]. Later in 1999, Yoon et al. fabricated an array of FeFETs based on ferroelectric SrBi$_2$Ta$_2$O$_9$ films, the array is operated as an electrically modifiable synapse circuit by carrying out the weighted sum operation[78]. In 2014, Kaneko et al. demonstrated continuous channel conductance change of ZnO channel by modulating the amplitude of gate voltage pulses in Pb(Zr,Ti)O$_3$ based FeFET[79, 89]. Assisted by a CMOS selector circuit, where a spike with positive and negative value is selectively applied to the gate of the FeFET, STDP learning rule is realized. Furthermore, on-chip pattern recognition of 3 × 3 matrix patterns is successfully demonstrated in a Hopfield neural network circuit consisting of CMOS neurons and FeFET synapses. In the unsupervised learning process, the conductance of all FeFET synapses was automatically and simultaneously adjusted in accordance with the STDP learning rule[79]. Note that this is the first time to report on-chip learning in memristor based hardware neural network.

CMOS-compatible HZO FeFETs are also widely researched as synaptic devices. Kim et al fabricated HZO FeFETs with indium gallium zinc oxide (IGZO) channel (Figure 13a). The IGZO conductance shows a clear hysteresis under the gate voltage (Figure 13b). By tuning polarization changes in the nanoscale ferroelectric layer, the IGZO conductance shows linear potentiation and depression characteristics with 65 multiple states in a range of 14.4 (Figure 13c). The simulations of a two-layer perceptron neural network with these HZO FeFETs achieves 91.1% recognition accuracy of handwritten digits. Jerry et al. demonstrated a HZO based FeFET synapse with 32 tunable conductance states and 75ns update rate[80]. These FeFET synapses show extremely symmetric potentiation and depression characteristics. The simulation implies that FeFET synapses show a $10^3$ to $10^6$ acceleration in online learning latency over multi-state RRAM based analog synapses[80]. These demonstrations suggest that FeFET synapses are promising for realizing the neural network hardware.

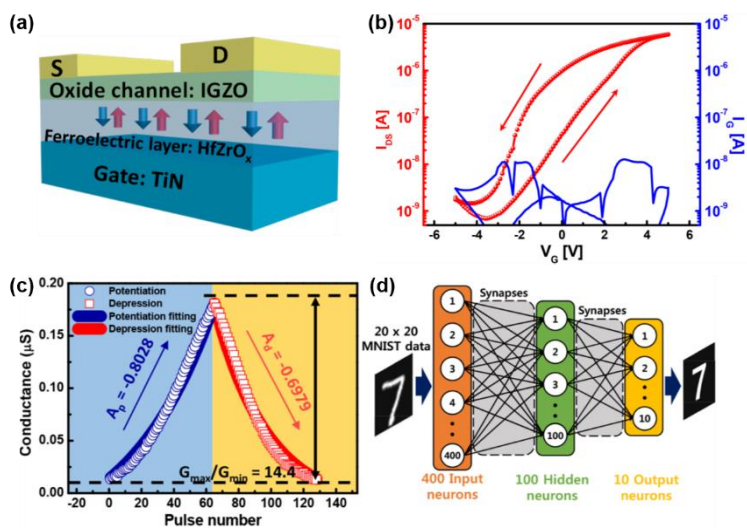

Figure 13 (a) Schematic structure of an HZO FeFET. (b) Transfer curve of the HZO FeFET. (c) LTP/D properties of the HZO FeFET with incremental pulse scheme. (d) Schematic illustration of two-layer perceptron neural network for simulation of pattern recognition. Reproduced from Ref. [83], Copyright 2019, with permission of ACS.

**4.2 Organic FeFET synapses**

Recently, organic FeFET, taking advantage of flexible device integration and large-scale fabrication, also attract much attention to be a synapse device[1, 84, 85, 87]. As presented in table 1, synaptic functions like LTP/D, EPSC/IPSC and STDP are demonstrated in single organic FeFET device. Assisted by organic FeFET synapses, visual-perception[84] and simulation of pattern classification[85] or image recognition[87] are realized in neural networks.

Specially, organic ferroelectric polymers and 2D semiconductors could be well combined. The ferroelectric domains could effectively modulate carrier density in 2D semiconductors[90]. Tian et al. fabricated an organic FeFET combining ferroelectric polymers and 2D semiconductors (Figure 14b)[1]. By altering the voltage amplitude applied on the gate electrode, the conductance of $MoS_2$ channel shows multi progressive evolution loops (Figure 14c) with a conductance range of $10^4$, in which the resistive threshold voltage corresponds well with the coercive voltage of ferroelectric gate polymer (Figure 14d). The in situ real-time correlation between dynamic resistive switching and polarization change is experimentally confirmed using transient current measurements. The conductance of this $MoS_2$ FeFET can be precisely manipulated to vary between more than 1000 intermediate states. Synaptic functions such as LTP/D (Figure 14e) and Hebbian STDP learning rule (Figure 14f) are demonstrated in these organic FeFET synapses. In addition, the device is expected to experience $1 \times 10^9$ synaptic spikes with an ultralow energy consumption for each synaptic operation (less than 1 fJ, compatible with a bio-synaptic event)[1], which highlights its immense potential for the massive neural architecture in bioinspired networks.

Using graphene, whose Dirac point could be shifted by polarization[91], instead of $MoS_2$ channel, Chen et al. demonstrated a complementary synapse using two graphene FeFET devices with opposite polarization direction[85]. As shown in Figures 15a-b, when the polarization is upward/downward, the graphene channel becomes hole/electron dominated due to different positions of Fermi levels within the graphene energy bands. Applying same pulse sequences for fine tuning of polarization, the channel conductance shows opposite variation between devices with polarization upward and downward, respectively. Remote supervise method (ReSuMe) circuits composed of graphene FeFET complementary synapses and leaky integrate-and-fire (LIF) neuron are designed (Figures 15c). A potentiative graphene FeFET and a depressive graphene FeFET are connected parallelly as complementary synapses. Their source and drain terminals are for receiving spikes from the presynaptic neuron and transmitting them to the postsynaptic neuron, respectively. The gate terminal of potentiative one and depressive one is for receiving supervised pulse only triggered by teach signal and output signal, respectively, to adjust their channel conductance. The amplitude of supervised voltage decays with time, guaranteeing the success of the supervised learning which is represented by the according timings between teach signal and output signal. For example, in this case of the first-round output fires earlier than the desired ($t_{out} < t_d$), the amplitude of supervised voltage applied on depressive one is greater than that applied on potentiative one ($c_1 > c_2 > 0$). As a result, the summing conductance of the complementary synapses decreases and consequently the second-round output fires later (Figures 15c). In this way the output timing approaches the desired one round-by-round. As shown in Figures 15d-e, pattern classification of 3x3 binary images of z, v, and n are realized in a single-layer SNN based on ReSuMe algorithm. In this ReSuMe learning of SNN, the key requirement on device properties is the symmetry of conductance tuning between the potentiative and depressive synapses. However, as seen in Figure 14e, symmetric conductance update in FeFET devices is challenging, and more efforts should be devoted to controlling mechanisms and rules of channel conductance by ferroelectric polarization.

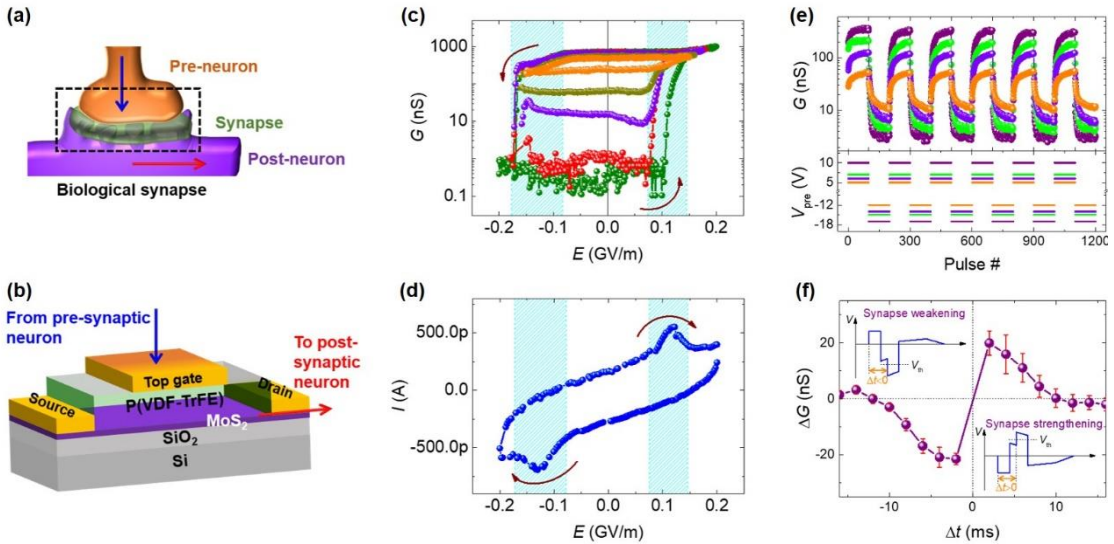

Figure 14 (a) Sketch of a biological synapse. (b) Sketch of an MoS$_2$ FeFET with ferroelectric P(VDF-TrFE) gate insulator. Note that the input port can be either the gate-source/drain port or the drain-source port. For surprised learning in general SNN, the channel usually connects pre and post synaptic neurons while the gate terminal accepts signals to modulate the channel conductance. (c) Channel conductance versus gate electric field curves in a MoS$_2$ FeFET. (d) Gate current versus gate electric field curve on the MoS$_2$ FeFET. (e) Evolution of channel conductance as a function of the different voltage pulse sequences. (f) Measurements of Hebbian STDP learning rule in a MoS$_2$ FeFET. Reproduced from Ref. [1], Copyright 2019, with permission of John Wiley and Sons.

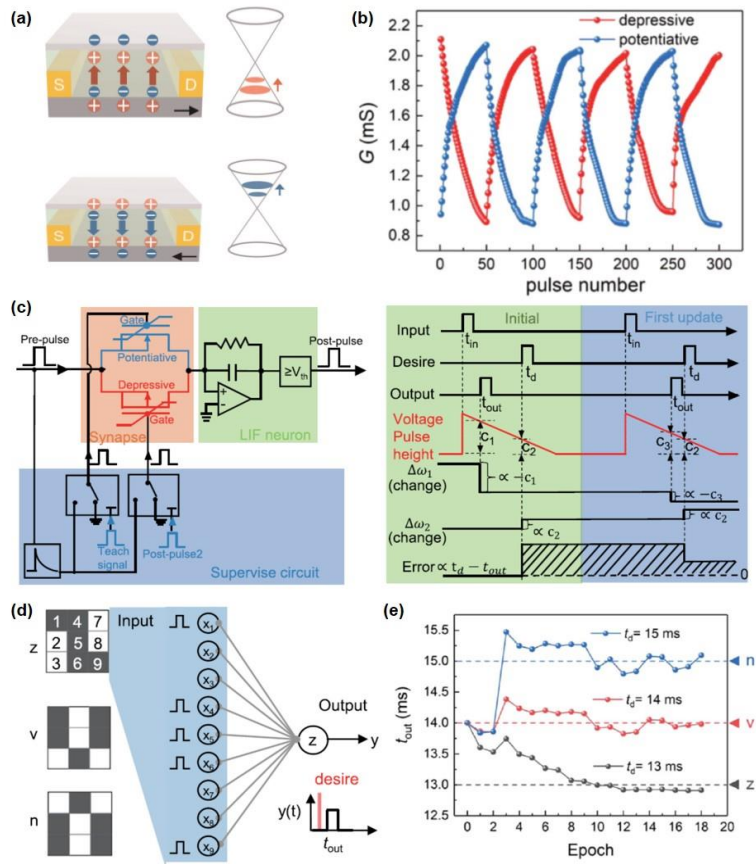

Figure 15 (a) Schematic view that when the polarization is upward/downward, the graphene channel becomes hole/electron dominated due to different positions of Fermi levels within the graphene energy bands. (b) The channel conductance will be decreased/increased due to reduction/enhancement of hole/electron density caused by the corresponding change of ferroelectric polarization. (c) A remote supervise method (ReSuMe) module composed of graphene FeFET complementary synapses, leaky integrate-and-fire (LIF) neuron, and supervise circuits (left). Time chart of signals (right). (d) Sketch of single-layer perceptron for classification of 3x3 binary images of z, v, and n, where classification is represented by the different timings of the output neuron. (e) The simulation results, the evolution of output signals, where output timings for z, v, and n inputs gradually approach are the desired timings labelled by dash lines. Reproduced from Ref. [85], Copyright 2019, with permission of Springer Nature.

## 5. Perspectives and challenges

The multiple ferroelectric polarization tuned by external electric field could be used to simulate the biological synaptic weight. The vector matrix multiplication (VMM), which is the basis for parallel computation in ANN, is usually realized by following the Kirchhoff's current law[59]. Thus, ferroelectric synaptic devices or "ferroelectric synapses" should transmit multiple polarization values to conductance states. This kind of ferroelectric synaptic devices includes two-terminal FTJs and FDs, and three-terminal FeFETs. It should be noted that the memory effect in ferroelectric capacitors is based on the destructive reading of charges, which is not appropriate for synapses. The weight is controlled by the input of the two-terminal weight cell whereas the weight is controlled by the gate that is independent of the input in a three-terminal cell structure.

These ferroelectric synapses have two advantages compared with other reported ones: one is the intrinsic switching of ferroelectric domains without invoking of defect migration as in resistive oxides contributes reliable performance in these ferroelectric synapses once reliable fabrication parameters are established. Another tremendous advantage is the extremely low energy consumption because the ferroelectric polarization is manipulated by electric field which eliminates the Joule heating by current as in magnetic and phase change memory. Ferroelectric synapses are potential for the construction of low-energy and effective brain-like intelligent networks.

However, despite these pioneering works summarised above, no large-scale hardware neural network based on ferroelectric synapses is demonstrated. Below are perspectives and some issues to be conquered for practical application of these ferroelectric synapses followed by several considerations for the application of ferroelectric synapses on the neural network level.

Table 2 Radar chart to summarize perspectives for the FTJs, FDs and FeFETs as electronic synapses.

|  | Ratio | States | Retention | Energy | Endurance | Concurrent learning | High density | Reported array |
|---|---|---|---|---|---|---|---|---|
| FTJs | good | good | good | poor | poor | no | good | yes |
| FDs | good | good | good | poor | poor | no | good | no |
| FeFETs | good | good | poor | good | good | yes | poor | yes |

FTJs are promising synaptic devices with large OFF/ON ratios[25], high uniformity[45] and high operating speed[49]. It should be mentioned the current situation that most of the experimentally reported FTJs are based on ferroelectric perovskite barriers. Sophisticated experimental approaches such as strain engineering and careful control of epitaxial growth are required for ferroelectricity in nanometer-thick perovskite films, which inevitably results in a scaling issue. The new emerging CMOS-compatible $HfO_2$-based ferroelectric films, fabricated by industry preferred magnetron sputtering or atomic layer deposition, and solution processed organic polymers, are very promising for the practical FTJ synapses. Another issue in FTJs is the endurance of only up to a few $10^8$ cycles[49], which is further inferior compared with the capacitive ferroelectric memories ($> 10^{16}$ cycles). The degeneration of ferroelectricity may result from the pinning effect of domain walls due to rearrangement of defects. This pinning effect is much stronger in nanometer-thick ferroelectric films of FTJs. One of solutions may be doping the ferroelectric material to reduce the mobility of defects.

FDs have the similar structure with FTJs but the ferroelectric barrier is much thicker, thus taking advantage of easy fabrication process and more stable performance properties. It is expected to realize synaptic functions by tuning the polarization of ferroelectric layer in FDs. However, defects located at the interface usually also play a role for resistive switching[77], which disturbs the whole dynamics. Efforts for controlling a high-quality interface in FDs are needed. Another issue is the contradiction between the ferroelectricity and conductivity of the ferroelectric layer. The conductivity contributes to the leakage of the FDs but degrades the ferroelectricity, while both conductivity and ferroelectricity are preferred by the resistive switching behaviour. Thus, ferroelectrics with narrow bandgaps are preferred and FDs using new emerging ferroelectrics with narrow bandgap should be explored. Doping techniques used in ferroelectric photovoltaics which effectively shrink the bandgap without significantly reducing the ferroelectricity are useful and could be directly applied for exploring well-performance FDs. Other solutions such as addition of metal nanoparticles to ferroelectric films are also reported to be successful[92-94].

FeFETs are three-terminal ferroelectric synapses, in which the input port can be either the gate-source/drain port or the drain-source port. When the drain/source port connects pre/post synaptic neuron and the gate terminal accepts signals to modulate the channel conductance by tuning ferroelectric polarization, FeFETs allow concurrent learning through direct feedback modulation to synapse weight from the gate electrode independently. On-chip pattern recognition in a hardware neural network based on $Pb(Zr,Ti)O_3$ FeFETs is successfully demonstrated. From the view of application, the thickness of ferroelectrics in FeFETs is desired to be minimized to reduce the operational voltage of ferroelectric polarization. It is faced with the same scaling issue of perovskite oxides as in FTJs. The $HfO_2$-based ferroelectric films demonstrate ferroelectricity at the thickness of one nanometer[61] and the incorporation of $HfO_2$-based ferroelectric films in field effect transistors is already well developed and implemented in CMOS technology. Efforts for optimizing the performance of $HfO_2$-based FeFETs are still needed. For example, the interfacial layer (IL) issue, Si reacting with $HfO_2$ to achieve the $SiO_2$ layer, not only increases operation voltage but also induces charge trapping, resulting in degradation of memory performance.

On the parallel, flexible ferroelectric synapses are also attracting, mainly due to their capacity of being on paper or plastics and wearable applications. The solution-processed ferroelectric polymers are extremely low-cost and easy for flexible device integration and large-scale fabrication. The bandgap of ferroelectric polymers is ~6 eV. The high insulativity makes it suitable for either tunneling barrier in FTJs and gate layer in FeFETs. These merits of low-loss and no leakage path in ferroelectric films are very important for 3D stacked synaptic weights, which could immensely reduce the neuromorphic chip size.

Several points for the application of ferroelectric synapses on the neural network level should also be emphasized.

**Sneak paths in ferroelectric neural network:** The tunnelling current intrinsically present non-linear relationship with applied electric field in FTJs, while the currents through the heterostructure of FDs show a rectifying characteristic. On one hand, both the nonlinearity in FTJs and self-rectification in FDs benefit to eliminate sneak paths in a passive crossbar array[59]. On the other hand, vector matrix multiplication (VMM), which is the basis for parallel computation in ANN, could be implemented in an analog manner by synapse crossbar arrays following $I_j = \sum_i V_i G_{ij}$ (current domain method) or $Q_j = \sum_i Vt_i G_{ij}$ (charge domain method). It is limited to charge domain method (by modulating pulse width or number) when performing VMM computation in FTJs or FDs crossbar arrays due to the nonlinear current versus voltage (*I-V*) curves. For FeFETs crossbar arrays, the controlling of third gate line makes the array free of sneak paths. Furthermore, a liner (*I-V*) relationship between source and drain of these FeFETs is expected, which is suitable to perform VMM with a current manner.

**Retention:** In the VMM process, the value of matrix cell is encoded as the synaptic weight (conductance) of the ANN. It asks a wide weight-bit and good resistance stability to reduce the error for VMM operation, which is very important for high accuracy of image processing. Thus, large remnant polarization ($P_r$) and good retention of intermediate states are preferred in ferroelectric synapses. The depolarization phenomenon in ferroelectric memories is a long talking story, and becomes more severe in ferroelectric synaptic devices. In ferroelectric synaptic devices, the polarization is transmitted to conductance. To enhance the dynamic range of conductance states, low charge screening elements are used, such as semiconductor electrode in FTJs and FDs and semiconductor channel in FeFETs, resulting in large depolarization field destabilizing the ferroelectric domains. Here we propose two aspects should be considered to alleviate the depolarization phenomenon. One is the control of high-quality interface to eliminate "dead layer" that plays the role of enlarging the depolarization field. For example, Ali et al. demonstrated that depositing a SiON interface between $HfO_2$ and Si effectively reduced the depolarization field and largely enhanced the retention behavior in $HfO_2$-based FeFETs[95]. The second is selecting ferroelectric films with coercive electric field as large as possible. For example, the coercive electric field in ultrathin ferroelectric polymer is ~ 0.2 GV/m, approaching the depolarization field even without any charge screening[37]. Covering ferroelectric polymers on 2D materials, ferroelectric domains with up and down direction could induce *p* and *n* regions in 2D semiconductors without degradation for three months[90].

**Uniformity**: Uniformity of cycle to cycle in one device and cycles of device to device are essential for performing VMM computation. Furthermore, the online training of the ANN requires linearity and symmetry of conductance update in synaptic devices for low energy consumption. To accomplish uniformities mentioned above in ferroelectric synapses, both ferroelectric domain dynamics and mechanisms underline resistive switching driven by ferroelectric polarization are needed to be further explored. The ferroelectric domain dynamics are reported to evolve from a domains progration model to homogeneous switching model as the thickness of ferroelectric films decreases from tens of nanometers to only few nanometers[51, 53, 96]. However, ferroelectric domains progration is observed in FTJs which may due to pinned defects at electrode considering the ultra large ratio between area and thickness ($> 10^4$)[38]. More experiments are needed to illustrate this problem.

For mechanisms underline resistive switching driven by ferroelectric polarization, only a direct correlation between ferroelectric domains and conductance states is further not enough. Indeed, large TER in FTJs cannot be quantitatively interpreted by simple electrostatic models and necessitates complex descriptions containing interfacial dielectric layers or doped-semiconducting layers[50]. The mechanism how does the polarization control the resistive switching behavior in FDs is still disputable and no direct experimental evidence for the interfacial depletion-layer width model is demonstrated. Finally, the nonlinear polarization against weighting (programming) voltage is also an issue. In a short, a rigid model, hybridizing both theory and experiments, guiding both the ferroelectric domain dynamics and resistive switching driven by ferroelectric polarization is urgently necessary, aiming at a liner and symmetric relationship between conductance and field stimulation (either amplitude or width).

**High density:** For a high-density ferroelectric neural network, the smallest lateral size of the ferroelectric film that maintains multiple properties is needed to be addressed. The control of multilevel of polarization for a small ferroelectric capacitor is challenging, which may need extremely short pulse width or small incremental voltage. It is reported that multilevel switching is difficult in nano-sized FeFETs due to either single grain or single domain[97]. Therefore, a stacked structure similar with vertical NAND may be induced for multibit weights of nanoscale ferroelectric synaptic devices.

Beyond the above challenges in ferroelectric synaptic devices like FTJs, FDs and FeFETs, some latterly emerging techniques, such as in-plane planar FTJs[98-100], Van Der Waals Ferroelectric[101], domain-wall memoy[102], and ferroelectric memtranstor[103], may provide new opportunities for application of ferroelectric synapses.

**Acknowledgement**

This work was financially supported by the National Natural Science Foundation of China (61804055), "Chenguang Progrom" supported by Shanghai Education Development Foundation and Shanghai Municipal Education Commission (17CG24), Shanghai Science and Technology Innovation Action Plan (19JC1416700).